\documentclass[conference]{IEEEtran}
\usepackage{graphicx,times,amsmath} 
\begin{document}

\title{Vector-Neuron Models of Associative Memory}
\author{\authorblockN{Boris V. Kryzhanovsky}
\authorblockA{Institute\\ of Optical Neural Technologies\\ Russian 
Academy 
of Sciences\\ Moscow, Russia\\
E-mail: kryzhanov@iont.ru}
\and
\authorblockN{Leonid B. Litinskii}
\authorblockA{Institute\\ of Optical Neural Technologies\\ Russian 
Academy 
of Sciences\\ Moscow, Russia\\
E-mail: litin@iont.ru}
\and
\authorblockN{Andrey L. Mikaelian}
\authorblockA{Institute\\ of Optical Neural Technologies\\ Russian 
Academy 
of Sciences\\ Moscow, Russia\\
E-mail: iont@postman.ru}}

\maketitle

\begin{abstract}
We consider two models of Hopfield-like associative memory with $q$-valued
neurons: Potts-glass neural network (PGNN) and parametrical neural network
(PNN). In these models neurons can be in more than two different states.
The models have the record characteristics of its storage capacity and 
noise immunity, and significantly exceed the Hopfield model.
We present a uniform formalism allowing us to describe both PNN and PGNN.
This networks inherent mechanisms, responsible for outstanding recognizing 
properties, are clarified. 
\end{abstract}

\section{Introduction}
The number of patterns that can be stored in Hopfield model (HM) is 
comparatively not so large. If $N$ is binary neurons number, then the 
thermodynamic approach leads to the well-known estimation of the HM storage 
capacity, $p_{HM}\sim 0.14\cdot N$ (\cite{Ami}, \cite{Her}).
At the early 90-th some authors suggested Hopfield-like models of 
associative memory with $q$-valued neurons that can be in more than two
different states, $q\ge 2$, \cite{Kan}-\cite{Bol4}.
All these models are related with the Potts model of magnetic. The 
last one generalizes the Ising model for the case of the spin variable that 
takes 
$q>2$ different values \cite{Wu},\cite{Bax}. In all these works the authors 
used 
the same well-known approach linking the Ising model with the Hopfield 
model 
(see, for example, \cite{Her}). Namely, in place of the short-range 
interaction 
between two nearest spins the Hebb type interconnections between all 
$q$-valued neurons 
were used. As a result, long-range interactions appear. Then in the 
mean-field 
approximation it was possible to calculate the statistical sum and, 
consequently, to construct the phase diagram. Different regions of the 
phase 
diagram were interpreted in the terms of the network ability to recognize 
noisy patterns.

For all these models, except one, the storage capacity is even less than 
that for HM. An exception is so named {\it  
Potts-glass neural network} (PGNN) \cite{Kan}. The numerical solution of 
transcendential equation system resulting from thermodynamic approach leads 
to 
the following estimation for storage capacity for PGNN
$$p_{PGNN}\sim\frac{q(q-1)}2\cdot p_{HM}.$$
As far as $q$-valued models are intended for color images processing, 
number  $q$  stands for number of different colors, used for elementary 
pixel 
can be painted.
Even if $q\sim 10$ the storage capacity of PGNN is 50 times as much as the 
storage capacity of HM. For computer processing of colored images the 
standard value is $q= 256$.
Consequently, comparing with HM the gain is about four orders,
$p_{PGNN}\sim 10^4\cdot p_{HM}$. 
It is very good result. However, for long time it was not clear, why PGNN 
has such a big storage capacity. Thermodynamic approach does not answer 
this 
question. 

On the other hand we worked out the model of associative memory, intended 
for implementation as an optical device (\cite{Kry02},\cite{Fon01}).  Such 
a 
network is capable to hold and handle 
information that 
is encoded in the form of the frequency-phase modulation. In the network 
the signals propagate along interconnections in the form of 
quasi-monochromatic pulses at $q$ different frequencies. 
There are arguments in favour of this idea. First of all, the 
frequency-phase 
modulation is more convenient for optical processing of signals. It 
allows us to back down an artificial adaptation of an optical network to 
amplitude modulated signals. Second, when signals with $q$ different 
frequencies can propagate along one interconnection this is an analog 
of the channel multiplexing. In fact, this allows us to reduce the 
number of interconnections by a factor of $q^2$.
Note that interconnections occupy nearly 98\% of the area of neurochips. 

In the center of our model the parametrical four-wave mixing process
(FWM) is situated, that is well-known in nonlinear optics \cite{Blo66}.
However, in order this model has good characteristics, an important 
condition must be added that should facilitate 
the propagation of useful signal, and, in the same time, 
suppress internal noise. This condition is {\it the principle of 
incommensurability of frequencies} proponed 
in \cite{Kry02},\cite{Fon01} in nonlinear optics terms (see Sec. 3).

The signal-noise analysis of our model made with the aid of the 
Chebyshev-Chernov statistical method \cite{Che52},\cite{Kry00} showed that 
the storage capacity of the network
was approximately $q^2$ times as much as the HM storage capacity.
We called our network {\it the parametrical neural network} (PNN).

We worked out {\it the vector formalism} -- universal description of PNN, 
not 
related directly to the optical model \cite{Kry4}-\cite{Kry6}. 
This formalism proved to be useful also for clear description of PGNN, although 
initially it was formulated in absolutely another terms. In this way one 
can 
easily establish relations between PGNN and PNN and also clarify the 
mechanisms, 
responsible for outstanding recognizing properties of both models.
The reason is the local architecture of both networks, which suppresses 
system internal noise. In other $q$-valued models there is no such 
suppression. 

In this paper we give PGNN description, using the vector formalism. Then we 
define our PNN, using nonlinear optics terms and the vector formalism as 
well. 
Moreover, we consider some possible architectures for PNN. 

{\bf Note.} Our vector formalism is almost identical to the vector-neuron 
approach, which was suggested some years ago by \cite{Nak}. We have found 
this 
paper after working out our own vector formalism. Dynamical rule in 
\cite{Nak} 
was formulated not in the best way, however it seems, that the authors of 
\cite{Nak} were the first to suggest the fruitful idea about representation 
of 
interconnections matrix as tensor product of vector-neurons.

\section{Potts-glass neural network}

We describe PGNN in terms of our vector formalism and in future compare it 
with PNN.
\subsection{Vector formalism}

PGNN consists of $N$ neurons each of which can be in $q$ different states. 
In order to describe the $q$ different states of neurons we use the set of
of $q$-dimensional vectors of a special type, so named {\it Potts 
vectors}.
Namely, the $l$th state of a neuron is described by a column-vector 
${\bf d}_l\in{\rm R^q}$,$${\bf d}_l=\frac1q\left(\begin{array}{c}-
1\\\vdots\\q-
1\\
\vdots\\-1\end{array}\right),\ l=1,\ldots,q.$$
The state of the $i$-th neuron is described 
by a vector 
${\bf x}_i={\bf d}_{l_i}, 1\le l_i\le q$.
The state of the network as a whole $X$ is determined by a set of $N$ 
column-vectors ${\bf x}_i$: $X=({\bf x}_1,\ldots,{\bf x}_N)$.
The $p$ stored patterns are 
$$\begin{array}{l}X^{(\mu)}=({\bf x}^{(\mu)}_1,\ldots,
{\bf x}^{(\mu)}_N),\ {\bf x}^{(\mu)}_i={\bf d}_{l^{(\mu)}_i},\\ 
1\le l^{(\mu)}_i\le q,\ \mu=1,2,\ldots,p.\end{array}$$ 

Since neurons are vectors, the local field 
${\bf h}_i$ affecting the $i$th neuron is a vector too, 
$${\bf h}_i  = \frac1N\sum\limits_{j = 1}^N {\bf T}_{ij}{\bf x}_j.$$
The $(q \times q)$-matrices ${\bf T}_{ij}$ describe the interconnections 
between the $i$th and the $j$th neurons. By analogy with the Hopfield model 
these matrices are chosen in generalized 
Hebb form:
$${\bf T}_{ij}=(1-\delta_{ij})\sum_{\mu=1}^p{\bf x}^{(\mu)}_i
{{\bf x}_j^{(\mu)}}^+,\quad i,j=1,\ldots,N,\eqno(1)$$
where ${\bf x}^+$ is $q$-dimensional row-vector and $\delta_{ij}$ is the 
Kronecker symbol. The matrix ${\bf T}_{ij}$ affects the vector 
${\bf x}_j\in\rm R^q$, converting it in a linear combination of  
column-vectors ${\bf d}_l$. After summation over all $j$ we get the local 
field ${\bf h}_i$ as linear combination of vectors ${\bf d}_l$ 
$${\bf h}_i=\sum_{l=1}^q A^{(i)}_l
{\bf d}_l.$$
Let $k$ be the index relating to the maximal coefficient: 
$A^{(i)}_k> A^{(i)}_l\ \forall\ l$. Then, by definition,
the $i$-th neuron at the next time step, $t+1$, is oriented 
along a direction mostly close to the local field ${\bf h}_i$ at the 
time 
$t$:
$${\bf x}_i(t+1)={\bf d}_k.\eqno(2)$$

The evolution of the system consists of consequent changes of orientations 
of vector-neurons according to the rule (2). We make the convention that if 
some of the coefficients $A^{(i)}_l$ are maximal simultaneously, and the 
neuron is 
in one of these {\it unimprovable} states, its state does not change. 
Then it is easy to show that during the evolution of the network its
{\it energy} $H(t)=-1/2\sum_{i=1}^N({\bf h}_i(t){\bf x}_i(t))$ decreases.
In the end the system reaches a local energy minimum. In this 
state all the neurons ${\bf x}_i$ are oriented in an unimprovable 
manner, and the evolution of the system come to its end. These states 
are the fixed points of the system. The necessary and sufficient 
conditions for a configuration $X$ to be a fixed point is fulfillment of 
the set of inequalities:
$$({\bf x}_i{\bf h}_i)\ge ({\bf d}_l{\bf h}_i),\quad 
\forall\ l=1,\ldots,q;\ \forall\ i=1,\ldots,N.\eqno(3)$$
When $q=2$, PGNN is the same as the standard Hopfield model. 

\subsection{Storage capacity of PGNN}
Let we have the randomized patterns $\{X^{(\mu)}\}_1^p$.
Suppose that the network starts from a distorted $m$th pattern 
$$\tilde X^{(m)}=(\hat b_1{\bf x}^{(m)}_1,\hat b_2{\bf x}_2^{(m)},
\ldots,\hat b_N{\bf x}_N^{(m)}).$$
The noise operator $\hat b_j$ with the probability $b$ 
changes the state of the vector ${\bf x}^{(m)}_j$, and with the probability 
$1-b$ this vector remains unchanged. 
In other words, $b$ is the probability of an 
error in a state of a neuron.
The noise operators $\hat b_j$ are 
independent, too. 

The network recognizes the reference pattern 
$X^m$ correctly, if the output of the $i$th neuron defined by 
Eq.(2) is equal 
to $\vec x_i^{(m)}$. Otherwise, PGNN fails to recognize the pattern $X^m$.
Let us estimate the probability of error in the recognition of $m$th 
pattern.

Simple calculations show, that probability of inequality validity $({\bf 
x}^{(m)}_i{\bf h}_i)<({\bf d}_l{\bf h}_i)$ at 
${\bf d}_l\ne{\bf x}^{(m)}_i$ can be expressed as
$${\rm Prob}\left\{\xi<\eta\right\}=
{\rm Prob}\left\{\frac1N\sum_{j\ne i}^N\xi_j<
\frac1N\sum_{j\ne i}^N\sum_{\mu\ne m}^p\eta_j^{(\mu)}\right\},\eqno(4)$$ 
where
$\eta_j^{(\mu)}=
({\bf d}_l-{\bf x}_i^{(m)},
{\bf x}_i^{(\mu)})
({\bf x}^{(\mu)}_j\hat b_j{\bf x}^{(m)}_j)$, $\xi_j=
({\bf x}^{(m)}_j\hat b_j{\bf x}^{(m)}_j)$.

The quantity $\xi$ is {\it the useful signal}. It is connected with 
influence of exactly the $m$th pattern onto the $i$th neuron. The 
partial random 
variables $\xi_j$ are independent and identically distributed. The 
quantity 
$\eta$ 
symbolizes {\it the inner noise}, connected with distorting influence of 
all 
other patterns. Partial noise components $\eta_j^{(\mu)}$ are independent 
and 
identically distributed. It is easy to obtain the distributions for
$\xi_j$ and $\eta_j^{(\mu)}$:
$$\left\{\begin{array}{ccl}\xi_j&=&\left\{\begin{array}{rl}(q-1)/q,&1-b\\-
1/q,&b\end{array}\right.,
\\ 
\\
\eta_j^{(\mu)}&=&\left\{\begin{array}{rl}(q-1)/q,&1/q^2\\
1/q,&(q-1)/q^2\\
0,&(q-2)/q\\
-1/q,&(q-1)/q^2\\
-(q-1)/q,&1/q^2\end{array}\right..\end{array}\right.\eqno(5)$$
Let us pay attention on the fact, that at $q>>1$ 
the noise component $\eta_j^{(\mu)}$ is localized mainly in zero:
$${\rm Prob}\left\{\eta_j^{(\mu)}=0\right\}=(q-2)/q\sim 1.$$

Total random variables $\xi$, $\eta$ are asymptotic normal distributed 
with 
parameters 
$$\begin{array}{ll}E(\xi)=\frac{q-1}q-b,&E(\eta)=0, \\
D(\xi)\to 0;& D(\eta)=\frac{2(q-1)}{q^3}\cdot\alpha.\end{array}\eqno(6)$$
where as usual {\it the loading parameter} $\alpha=\frac{p}N$. Now the 
probability 
of recognition error of coordinate ${\bf x}^{(m)}_i$ can be calculated by 
integration of the area under the "tail" of normally distributed $\eta$, 
where 
$\eta>E(\xi)$. 
Here we can explain, why the storage capacity of PGNN is much larger than 
HM.

The same considerations we can are valid for HM. It is done for example in 
\cite{Her}. Again we obtain a useful signal $\xi$  and an internal noise $\eta$, 
and Eq. (4) for the probability of 
recognition failure. Again these random 
quantities will asymptotic normal as sums of independent, identically 
distributed partial random components $\xi_j$ and $\eta_j^{(\mu)}$. The 
distributions of these last components can be obtained from Eq.(5) at $q=2$ 
(because PGNN transforms into HM in this case).  Mean values and dispersions for 
$\xi$ and $\eta$ can be obtained from (6) in the same way. As the result we 
have for HM:
$$\begin{array}{ll}
\xi_j=\left\{\begin{array}{rl}1/2,&1-b\\-1/2,&b\end{array},\right.&
\eta_j^{(\mu)}=\left\{\begin{array}{rl}1/2,&1/2\\
-1/2,&1/2\end{array},\right.\\
E(\xi)=\frac12-b,& E(\eta)=0,\\
D(\xi)\to 0;& D(\eta)=\frac\alpha{4}.\end{array}\eqno(7)$$
Comparison of (7) with (5) and (6) demonstrates, that the dispersion of 
internal noise for PGNN is much smaller, than that for HM: 
$$D_{PGNN}(\eta)/D_{HM}(\eta)=\frac{8(q-1)}{q^3}<<1, \mbox{ when $q>>1$.}
$$ 
Already at $q\sim 10$ the internal noise dispersion for PGNN is an order of 
magnitude smaller, than that for HM. Moreover, at $q\sim 10^2$ the fall of the 
dispersion is four orders of magnitude! This defines PGNN superiority over HM. 
We will give explanation of mechanism of internal noise compression in PGNN 
in the following Section. 

Switching from one vector-coordinate situation to that with the whole 
pattern and using the standard approximation (\cite{Kry5},\cite{Kry6}) 
we obtain the expression for the 
probability of the error in the recognition of the pattern $X^{(m)}$, 
$${\Pr}_{err}\sim\sqrt{Np}\exp\left(-\frac{N}{2p}\frac{q(q-1)}2(1-\bar 
b)^2\right),
\quad\bar b=\frac{q}{q-1}b.\eqno(8)$$ 
The expression sets the upper limit for the probability of 
recognition failure
for PGNN. Then, the asymptotically possible value of the storage capacity 
of 
PGNN is
$$p_c=\frac{N}{2\ln N}\frac{q(q-1)}2(1-\bar b)^2.\eqno(9)$$ 
When $q=2$, these expressions give the known estimates for HM.
For $q>2$ the storage capacity of PGNN is $q(q-1)/2$ times as large as the 
storage capacity of HM. In \cite{Kan} the same factor was 
obtained by fitting the results of numerical calculations. We
obtain the same result rigorously.

\section{Parametrical neural network}
Here we describe our associative memory model both in nonlinear optics and 
vector-formalism terms. We also will set out the obtained results for this 
model. 
\subsection{Nonlinear optic formulation}
In the network 
the signals propagate along interconnections in the form of 
quasi-monochromatic pulses at $q$ different frequencies 

$$\{ \omega _l \}_1^q  \equiv \{ \omega _1 ,\omega _2 ,...,\omega _q \}.
\eqno(10)$$
The model is based on a parametrical neuron that is a cubic nonlinear 
element capable to transform and generate frequencies in the parametrical 
FWM-processes
$\omega_i  - \omega _j  + \omega _k  \to \omega _r$.
Schematically this model of a neuron can be assumed as a device that is
composed of a summator of input signals, a set of  $q$ ideal frequency 
filters
$\{ \omega _l \}^q$, a block comparing the amplitudes of the signals 
and $q$ generators of quasi-monochromatic signals $\{ \omega _l \} ^q$.

Let $\{K^{(\mu)}\}_1^p$ be a set of patterns each of which is a set of 
quasi-monochromatic pulses with frequencies defined by Eq.(10) and 
amplitudes equal to
$\pm 1$: 
$$\begin{array}{l}K^{(\mu)}=(\kappa_1^{(\mu)},\ldots,\kappa_N^{(\mu)}),
\ 
\kappa_i^{(\mu)}=\pm\exp(\imath\omega_{l_i^{(\mu)}}t),\\
\mu=1,\ldots,p;\ i=1,\ldots,N;\ 1\le l_i^{(\mu)}\le 
q.\end{array}\eqno(11)$$
The memory of the network is localized in interconnections $T_{ij},\ 
i,j=1,\ldots,N$, which accumulate the information about the states of
$i$th and $j$th neurons in all the $p$ patterns. We suppose that the 
interconnections are dynamic ones and that they are organized according to
the Hebb rule:
$$T_{ij}=(1-\delta_{ij})\sum_{\mu=1}^p\kappa_i^{(\mu)}\kappa_j^{(\mu)*},
\quad i,j=1,\ldots,N.\eqno(12)$$
The network operates as follows. A quasi-monochromatic pulse with a 
frequency
$\omega_{l_j}$ that is propagating along the $(ij)$-th interconnection 
from the
$j$th neuron to the $i$th one, takes part in FWM-processes with the pulses 
stored 
in the interconnection,
$\omega_{l_i^{(\mu)}}-
\omega_{l_j^{(\mu)}}+\omega_{l_j}\to\{\omega_l\}_1^q$.
The amplitudes $\pm 1$ have to be multiplied. Summing up the results of 
these partial transformations over all patterns, $\mu=1,\ldots,p$, we 
obtain a 
packet 
of quasi-monochromatic pulses, where all the frequencies from the set 
(10) are present.
This 
packet is the result of transformation of the pulse $\omega_{l_j}$ by 
the interconnection $T_{ij}$, and it comes to the $i$th neuron. All such 
packets are summarized in this neuron. The summarized signal propagates 
through $q$ parallel ideal frequency filters. The output signals from the 
filters are compared with respect to their amplitudes. The signal with 
the maximal amplitude activates the $i$-th neuron  ('winner-take-all'). As 
a 
result it 
generates an output signal whose frequency and phase are the same as the 
frequency and the phase of the activating signal. 

Generally, when three pulses interact, under a FWM-process always the 
fourth pulse appears. The frequency of this pulse is defined by the 
conservation laws only. However, in order that the abovementioned model 
works 
as a memory, an important condition must be add, which has to facilitate 
the propagation of the useful signal, and, in the same time, to 
suppress external noise. This condition is {\it the principle of 
incommensurability of frequencies} proponed in \cite{Kry02},\cite{Fon01}: 
{\it no combinations $\omega_l-\omega_{l'}+\omega_{l''}$ can belong to the 
set 
(10), when all the frequencies are different.}

Now we finished to describe the principle of the network operating.
This network will be called {\it the parametrical neural network} (PNN).
Here an 
important remark has to be done. 

Generally speaking, there are different 
parametrical FWM-processes complying with the principle of 
incommensurability of 
frequencies. However,
better results can be obtained for the parametrical FWM-process  
$$\omega_l-\omega_{l'}+\omega_{l''}=\left\{
\begin{array}{cl}
\omega_l,&
\mbox{ when }l'=l'';\\
\rightarrow 0,&\mbox{in other cases.}\end{array}
\right.\eqno(13)$$
This architecture will be called PNN-2 (another 
architecture, PNN-1, was examined in \cite{Kry02},\cite{Fon01}).
Here we investigate the abilities of PNN-2. 
The structure of the rest of the paper is as follows. In 
next subsection we introduce a 
vector formalism allowing us to formulate the problem in the general 
form. Then, the results for PNN-2 will be presented. 
Then we mention shortly about other neuro-architectures, based on PNN-2. 
Some remarks are given in Conclusions.

\subsection{Vector formalism for PNN-2}

In order to describe the $q$ different states (10) of neurons we use the 
set of basis vectors ${\bf e}_l$ in the space $\rm R^q,\ q\ge 1$,
$${\bf e}_l=\left(\begin{array}{c}0\\\vdots\\1\\
\vdots\\0\end{array}\right),\ l=1,\ldots,q.$$
The state of the $i$th neuron is described by a vector 
${\bf x}_i$,
$${\bf x}_i=x_i{\bf e}_{l_i},\ x_i=\pm 1,\ {\bf e}_{l_i}\in{\rm 
R^q},\left\{\begin{array}{l} 1\le l_i\le q;\\ i=1,\ldots,N.\end{array}
\right.\eqno(14)$$ 
The factor $x_i$ denotes the signal phase. 
The state of the network as a whole $X$ is determined by a set of $N$ 
$q$-dimensional vectors ${\bf x}_i$: $X=({\bf x}_1,\ldots,{\bf x}_N)$.
The $p$ stored patterns are 
$$\begin{array}{l}X^{(\mu)}=({\bf x}^{(\mu)}_1,{\bf x}^{(\mu)}_2,\ldots,
{\bf x}^{(\mu)}_N),\ 
{\bf x}^{(\mu)}_i=x^{(\mu)}_i{\bf e}_{l^{(\mu)}_i},\\ 
x^{(\mu)}_i=\pm 1,\ 1\le l^{(\mu)}_i\le q,\ \mu=1,\ldots,p,\end{array}$$ 
and the local field is 
$${\bf h}_i  = \frac1N\sum\limits_{j = 1}^N {\bf T}_{ij}{\bf 
x}_j.\eqno(15)$$
The $(q \times q)$-matrix ${\bf T}_{ij}$ describes the interconnection 
between the $i$th and the $j$th neurons. This matrix affects the vector 
${\bf x}_j\in\rm R^q$, converting it in a linear combination of basis 
vectors ${\bf e}_l$. This combination is an analog of the packet of 
quasi-monochromatic pulses that come from the $j$th neuron to the $i$th 
one after transformation in the interconnection. To 
satisfy the conditions (12) and (13), we need to take 
the matrices ${\bf T}_{ij}$ as
$${\bf T}_{ij}=(1-\delta_{ij})\sum_{\mu=1}^p{\bf x}^{(\mu)}_i
{{\bf x}_j^{(\mu)}}^+,\quad i,j=1,\ldots,N.\eqno(16)$$
Note, that the structure of this expression is similar to that of (1). 

The dynamic rule is left as earlier:
the $i$th neuron at the time $t+1$ is 
oriented along a direction mostly close to the local field ${\bf h}_i(t)$. 
However the expressions will differ from (2). Indeed,
with the aid of (16) we 
write ${\bf h}_i$ in the form more convenient for analysis:
$${\bf h}_i(t)=\sum_{l=1}^q A^{(i)}_l
{\bf e}_l,
\ 
A^{(i)}_l\sim\sum\limits_{j(\ne i)}^N\sum\limits_{\mu=1}^p({\bf e}_l
{\bf x}^{(\mu)}_i)({\bf x}^{(\mu)}_j{\bf x}_j(t)).\eqno(17)$$
Let $k$ be the index relating to the amplitude that is maximal {\it in 
modulus} in the series (17): 
$\mid A^{(i)}_k\mid>\mid A^{(i)}_l\mid \ \forall\ l$.
Then according to our definition,
$${\bf x}_i(t+1)={\rm sgn}(A^{(i)}_k){\bf e}_k.\eqno(18)$$
The evolution of the system consists of consequent changes of orientations 
of vector-neurons according to the rule (18). The necessary and sufficient 
conditions for a configuration $X$ to be a fixed point is fulfillment of 
the set of inequalities:
$$({\bf x}_i{\bf h}_i)\ge \mid ({\bf e}_l{\bf h}_i)\mid,\quad 
\forall\ l=1,\ldots,q;\ \forall\ i=1,\ldots,N,$$
(compare with Eq.(3)).

\subsection{Storage capacity of PNN-2}
All these considerations are identical to those for PGNN. Differences 
appear only because of neurons are defined now not only by vectors, but also by 
scalars $\pm 1$. 
The distorted $m$th pattern has the form
$$\tilde X^{(m)}=(a_1\hat b_1{\bf x}^{(m)}_1, a_2\hat b_2{\bf 
x}_2^{(m)},
\ldots,a_N\hat b_N{\bf x}_N^{(m)}).$$
Here $\{a_i\}_1^N$ and $\{\hat b_i\}_1^N$ define a {\it phase noise} and a 
{\it frequency noise}
respectively: $a_i$ is a random value that is equal to 
$-1$ or $+1$ with the probabilities $a$ and $1-a$ respectively; $b$ is 
the probability that the operator $\hat b_i$ changes the state of the 
vector ${\bf x}^{(m)}_i=x^{(m)}_i{\bf e}_{l^{(m)}_i}$, and $1 - b$ 
is the probability that this vector remains unchanged.

The amplitudes $A^{(i)}_l$ (17) have the form 
$$A^{(i)}_l\sim\left\{\begin{array}{rcll}
x^{(m)}_i\sum_{j\ne i}^N\xi_j&+&
\sum_{j\ne i}^N\sum_{\mu\ne m}^p\eta_j^{(\mu)},&l=l^{(m)}_i;\\
\\
{}&{}&\sum_{j\ne i}^N\sum_{\mu\ne m}^p\eta_j^{(\mu)},&l\ne 
l^{(m)}_i,
\end{array}\right.$$
where $\eta_j^{(\mu)}=a_j({\bf e}_l{\bf x}^{(\mu)}_i)
({\bf x}^{(\mu)}_j\hat b_j{\bf x}^{(m)}_j)$, 
$\xi_j=a_j({\bf x}^{(m)}_j\hat b_j{\bf x}^{(m)}_j)$, 
 $j(\ne i)=1,\ldots,N$,
$\mu(\ne m)=1,\ldots,p$. 
When the patterns  
$\{X^{(\mu)}\}_1^p$ are uncorrelated, the quantities  
$\xi_j$ and $\eta_j^{(\mu)}$ are independent random variables 
described by the probability distributions
$$\xi_j=\left\{ {\begin{array}{*{20}c}
   { + 1,}  \\
   {0,}  \\
   { - 1}  \\
\end{array}} \right.\begin{array}{*{20}c}
   {(1 - b)(1 - a)} \hfill  \\
   b \hfill  \\
   {(1 - b)a} \hfill  \\
\end{array}
,\ 
\eta_j^{(\mu)}=\left\{{\begin{array}{*{20}c}
   { + 1,}  \\
   {0,}  \\
   { - 1}  \\
\end{array}} \right.\begin{array}{*{20}c}
   {1/2q^2 } \hfill  \\
   {1 - 1/q^2 } \hfill  \\
   {1/2q^2 } \hfill  \\
\end{array},
$$
(compare with Eq.(5)). As in the case of PGNN, when $q>>1$ the noise 
component $\eta_j^{(\mu)}$ is localized mainly in zero:
$${\rm Prob}\left\{\eta_j^{(\mu)}=0\right\}=1-1/q^2\sim 1.$$
Eq.(6) now will transform into:
$$\begin{array}{ll}E(\xi)=(1-2a)(1-b),&E(\eta)=0, \\
D(\xi)\to 0;& D(\eta)=\frac1{q^2}\cdot\alpha.\end{array}$$
When $q>>1$ the dispersion of internal noise for PNN-2 is even smaller, 
than for PGNN:
$$D_{PNN}(\eta)/D_{PGNN}(\eta)=1/2, \mbox{ when $q>>1$.}$$ 
In the long run this determines the superiority of PNN-2 over PGNN in 
memory capacity and noise immunity.  
It is convenient here to mention mechanisms, suppressing internal noises. 
They are identical in both models, but we will demonstrate them on the PNN 
example.  

When signal propagates it interacts with frequencies, stored in 
interconnection 
$\omega_{l_i^{(\mu)}}-\omega_{l_j^{(\mu)}}+\omega_{l_j}\rightarrow 
\{\omega_l\}_1^q$.
In addition the principal of frequencies incommensurability (13) should be 
fulfilled. 
It can be formulated in vector terms as: 
$${\bf x}^{(\mu)}_i{{\bf 
x}_j^{(\mu)}}^+{\bf x}_j=
\left\{
\begin{array}{rl}
{\bf x}^{(\mu)}_i,&
\mbox{when }l_j^{(\mu)}=l_j;\\
0,&\mbox{in other cases.}\end{array}\right.$$
One can see from the last equation, that the largest part of propagated 
signals will be suppressed. It happens because the interconnection chooses the 
only one combinations of indices $l_j^{(\mu)}$ and $l_j$ from all possible ones, 
where indices coincide (other combinations give zero). 
In other words, the interconnection filters signals. It is the main reason 
of the largest part of internal noise $\eta$ is localized in zero. 

The similar filtration happens also in PGNN. The difference is that in PGNN 
the signal always propagates through the interconnection. But when indices 
$l_j^{(\mu)}$ and $l_j$ coincide, the signal is attributed with large positive 
amplitude $\sim 1$. If indices do not coincide, the signal is attributed with 
small negative amplitude $\sim -1/q$. 
This signal filtration leads to suppression of internal noise in PGNN. In 
all another $q$-valued models of associative memory this filtration is absent.

At the end of consideration of PNN-2 we give the expressions for noise 
immunity and storage capacity similar to (8) and (9):

$${\Pr}_{err}\sim\sqrt{Np}\exp\left(-\frac{N(1 - 2a)^2}{2p}\cdot q^2(1-
b)^2\right),\eqno(19)$$
$$p_c=\frac{N(1-2a)^2}{2\ln N}\cdot q^2(1 - b)^2.\eqno(20)$$

When $q = 1$, Eqs.(19)-(20) transform into 
well-known results for the standard Hopfield model
(in this case there is no frequency noise, $b=0$). When $q$ increases, 
the probability of the error (19) decreases exponentially, i.e. the
noise immunity of PNN increases noticeably. In the same time the storage 
capacity of the network increases proportionally 
to $q^2$. In contrast to the Hopfield model the number of the 
patterns $p$ can be much greater than the number of neurons. 

For example, let us set a constant value ${\Pr}_{err}=0.01$. In the 
Hopfield 
model, with this probability of the error we can recognize any of $p=N/10$ 
patterns, each of which is less then 30\% noisy. In the same time,  PNN-2 
with 
$q=64$ allows us to recognize any of $p=5N$ patterns with 90\% noise, or 
any of $p=50N$ patterns with 65\% noise. Our computer simulations confirm 
these results.

The memory capacity in PNN-2 is twice as large as that in PGNN. Evidently, 
it is connected with the fact, that for the same $q$ the number of different 
states of neurons in PNN-2 is twice as large as that in PGNN. 
 In general, both models have very similar characteristics. 
\subsection{Other PNN-architectures}

\subsubsection{Phase-independent PNN-3}
When the PNN is realized as a device, the problem arises, that one should 
control the phases of all signals. All phases should be matched. It is rather 
difficult problem. 
It seems, that the easiest way to overcome this difficulty is to make all 
phases identical. Formally, we should make all amplitudes $\pm 1$ in (11) and 
(14) to $1$. More precise analysis shows, that in this case partial noise 
components $\eta_j^{(\mu)}$ become not independent. The noise dispersion 
drastically increases.
The way out is to use specially chosen vector thresholds in the local field 
definition \cite{Ali}:
$${\bf h}_i  = \frac1N\left\{\sum\limits_{j = 1}^N {\bf T}_{ij}{\bf x}_j
-\frac1q\sum_{\mu=1}^p{\bf x}_i^{(\mu)}\right\},\eqno(21)$$
where matrices ${\bf T}_{ij}$ are determined by Eq.(16).
Then the partial noise components $\eta_j^{(\mu)}$ become uncorrelated. And 
it is possible to apply the probability-theoretic approach for estimation of 
signal/noise ratio.

Means and dispersions of total random variables $\xi$ and $\eta$ are the 
same as in expressions (6). But whole phase-independent PNN  (we called it as 
PNN-3) is equivalent to PGNN. If to compare with PNN-3-model PGNN is too 
complicated. It is related with using the Potts vectors ${\bf d}_l$ instead of 
basis vectors ${\bf e}_l$.
Being realized as a computer algorithm PNN-3 works $q$ times quicker than 
PGNN. 

   \subsubsection{Decorrelating PNN}
   We  suggested  the method of sufficient enlarging  of  binary
associative  memory  with the help of PNN-architecture  for  the
case of correlation between patterns (\cite{Kry8},\cite{Kry7}).
As  it is known the memory capacity of Hopfield model falls down
drastically if there are correlations, so the only way out is so
named {\it sparse coding} \cite{Pal}-\cite{Fro}. Our method  is
an alternative to this approach.
   
   At the heart of our approach is one-to-one mapping of binary
patterns into internal representation, using vector-neurons of 
large dimension, $q>>1$. Then PNN is being constructed on the
basis of obtained vector-neuron patterns. The representation  has
the  following  properties: {\it i)} correlations between vector-neuron
patterns become negligible; {\it ii)} dimension $q$ of vector-neurons
increases {\it exponentially} as a function of mapping parameter.
The larger a dimension $q$ the better recognition properties
of PNN. The result of exponential increase of $q$ leads to 
the exponential increase of  binary  memory capacity.

The mapping of binary patterns into vector-neuron ones is
based on the very clear idea. This idea resembles the method, 
which was used previously in sparse coding (\cite{Pal3}), 
where due to a redundant coding it was possible to increase 
the storage capacity comparing with the Hopfield model. 
In the same time the noise immunity of the system was very low. 
In our case the redundancy of coding is absent,  
the storage capacity increases drastically, and the noise immunity  
is much greater. In future we plan to compare PNN with sparse coding in details.

\section{Conclusions}
   From  the  early  90th  the intensity  of  $q$-valued  neural
networks  researches  sharply decreased. Presumably  it  can  be
explained  by  absence of progress in development  of  effective
models   of  associative  memory.  Computer  algorithm  of   
PNN-architecture demonstrates, that we approach to those  magnitudes
of storage capacity  and  noise 
immunity  which could be of interest for practical applications.
Use of PNN-architectures seems to us very promising.

\section*{Acknowledgment}
The authors would like to thank Dr. Tatyana Lisovsky for her help.

This work was supported by the 
program "Intellectual Computer Systems" (the project 2.45)
and in part by Russian Basic Research Foundation (grants 
02-01-00457 and 03-01-00355)


\begin{thebibliography}{99}

\bibitem{Ami} D. Amit, H. Gutfreund, H. Sompolinsky, 
"Storing Infinite Numbers of Patterns in a Spin-Glass Model of Neural 
Networks,"
{\it Phys. Rev. Lett.}, vol. 55, pp. 1530-1533, 1985; 
"Information storage in neural networks with low levels of activity,"
{\it Phys. Rev. A}, vol. 35, pp. 2293-2303, 1987.

\bibitem{Her} J. Hertz, A. Krogh, R. Palmer, {\it Introduction to the 
Theory of Neural Computation}, NY: Addison-Wesley, 1991.

\bibitem{Kan} I. Kanter, "Potts-glass models of neural networks," 
{\it Physical Review A}, vol. 37, pp. 2739-2742, 1988.

\bibitem{Noe} J. Noest, "Discrete-state phasor neural networks," 
{\it Physal Review A}, vol. 38, pp. 2196-2199, 1988.

\bibitem{Coo} J. Cook, "The mean-field theory of a Q-state neural
network model," {\it Journal of Physics A}, vol. 22, pp. 2000-2012, 1989.

\bibitem{Rie} H. Rieger, "Storing an extensive number of grey-toned patterns
in a neural network using multistate neurons," {\it Journal of Physics A}, 
vol. 23, pp. L1273-L1279, 1990.

\bibitem{Bol1} D. Bolle, P. Dupont, J. van Mourik, "Stability
properties of Potts neural networks with biased patterns and low loading,"
{\it Journal of Physics A}, vol. 24, pp. 1065-1081, 1991.

\bibitem{Vog92} H. Vogt, A. Zippelius, "Invariant recognition in 
Potts glass neural networks," {\it Journal of Physics A}, vol. 25, 
pp. 2209-2226, 1992. 

\bibitem{Bol2} D. Bolle, P. Dupont, J. Huyghebaert, "Thermodynamics 
properties of the q-state Potts-glass neural network," {\it Physical Rewiew A},
vol. 45, pp. 4194-4197, 1992.

\bibitem{Bol4} D. Bolle, J. van Mourik,  
"Capacity of diluted multi-state neural networks,"
{\it Journal of Physics A}, vol. 27, pp. 1151-1162, 1994.

\bibitem{Wu} F.Y Wu, "The Potts model," {\it Review of Modern 
Physics}, vol. 54, pp. 235-268, 1982.

\bibitem{Bax} R. Baxter, {\it Exactly Solved Models in Statistical
Mechanics,} London: Academic Press, 1982.

\bibitem{Kry02} B.V. Kryzhanovsky, A.L. Mikaelian, "On recognition
ability of neuronet based on neurons with parametrical frequencies 
convertion," {\it Doklady Mathematics}, vol. 65(2), pp. 286-288, 2002.

\bibitem{Fon01} A. Fonarev, B.V. Kryzhanovsky et al., "Parametric
dynamic neural network recognition power," {\it Optical Memory \& Neural 
Networks}, vol. 10(4), pp. 31-48, 2001. 

\bibitem{Blo66} N. Bloembergen, {\it Nonlinear optics}, NY: Benjamin, 
1965.

\bibitem{Che52} N. Chernov, "A mesure of asymptotic efficiency
for tests of hypothesis based on the sum of observations," 
{\it Ann. Math. Statistics}, vol. 23, pp. 493-507, 1952.

\bibitem{Kry00} B.V. Kryzhanovsky, A.L. Mikaelian, V.N. Koshelev, et al.,
"On recognition error bound for associative Hopfield 
memory," {\it Optical Memory \& Neural Networks}, vol. 9(4),  
pp. 267-276, 2000.

\bibitem{Kry4} B.V. Kryzhanovsky, L.B. Litinskii, A. Fonarev, 
"Optical neural network based on the parametrical four-wave mixing 
process,"
In: Wang, L., Rajapakse, J.C, Fukushima, K., Lee, S.-Y., and Yao, 
X. (eds.):
{\it Proceedings of the 9th International Conference on Neural 
Information
Processing (ICONIP'02)}, Orchid Country Club, Singapore, vol. 4, pp. 
1704-1707, 2002.

\bibitem{Kry5} B.V. Kryzhanovsky, L.B. Litinskii, "Vector models of
associative memory," {\it Automaton and remote control}, vol. 64(11),
pp. 1782-1793, 2003.


\bibitem{Kry6}  A.L. Mikaelian, B.V. Kryzhanovsky, L.B. Litinskii, 
"Parametrical Neural Network," {\it Optical Memory \& Neural Networks}, 
vol. 12(3), pp. 227-236, 2003.

\bibitem{Nak} Y. Nakamura, K. Torii, T. Munaka, "Neural-network model 
composed of
multidimensional spin neurona," {\it Phys. Rev. B}, vol. 51, pp. 1538-1546, 
1995.

\bibitem{Ali} D.I. Alieva, B.V. Kryzhanovsky, "Phaseless parametrical neural 
net," 
{\it International Conference on Artificial Intelligent Systems
IEEE AIS'2003}, vol. 1, pp. 511-517, Moscow, Fizmatlit, 2003 (in russian).

\bibitem{Kry8} B.V. Kryzhanovsky, A.L. Mikaelian,  
"An associative memory capable of recognizing strongly correlated patterns,"
{\it Doklady Mathematics}, vol. 67(3), pp. 455-459, 2003.

\bibitem{Kry7} B. Kryzhanovsky, L. Litinskii, and A. Fonarev, 
"An effective associative memory for pattern recognition,"  
In: {\it Advances in 
Intelligent Data Analysis V. 5th International Symposium on Intelligent 
Data 
Analysis
IDA 2003}, pp. 179-186, Berlin: Springer, 2003.

\bibitem{Pal} G. Palm, "Memory capacity of local rules for synaptic 
modification," {\it Concepts in Neuroscience}, vol. 2, pp. 97-128, 1991.

\bibitem{Pal1} G. Palm, "On the information storage capacity of 
local learning rules," {\it Neural Computation}, vol. 4, pp. 703-711, 1992.

\bibitem{Pal2} G. Palm, F.T. Sommer, "Information capacity in 
recurrent 
McCulloch-Pitts networks with sparsely coded memory states," {\it Network}, 
vol. 3, pp. 1-10, 1992. 

\bibitem{Ami1} C.J. Perez-Vicente, D.J.  Amit, "Optimized network 
for 
sparsely coded patterns," {\it Journal of Physics A}, vol. 22, pp. 559-569,
1989. 

\bibitem{Fro} A.A. Frolov, I.P. Murav'ev, "Informational characteristics 
of 
neural networks capable of associative learning based on Hebbiean 
plasticity," {\it Network}, vol. 4, pp. 495-536, 1995. 

\bibitem{Pal3} G. Palm, 2003, private comunication.
\end{thebibliography}
\end{document}